\documentstyle[prl,aps,tabularx,psfig]{revtex}
\begin{document}

\title{
Inelasticity and intranuclear cascading in geometrical multichain model}
\draft
\author{T. Wibig}

\address{Experimental Physics Dept., University of \L odz,
Pomorska 149/153, 90-236 \L odz, Poland}

\date{\today}

\maketitle

\begin{abstract}
The inelasticity in nucleus-nucleus collisions at high energies
is calculated in the framework of geometrical multichain model.
The very fast increase of the inelasticity is found as a result of a 
second-stage 
cascading process. The same behaviour is expected for all the models
using the wounded nucleon idea. The simple formula for the number of
wounded nucleons inside colliding nuclei does not need to be exact.
\end{abstract}


\section {Introduction}
The inelasticity in nuclei collisions is one of the most important
properties of the interaction from the point of view of hadronic cascade
development in matter. Cosmic-ray extensive air shower (EAS) physics
is a particular domain where the knowledge of inelasticity is
extremely important.

The definition of inelasticity itself is a problem. The definition which is
a subject of this paper will be given
in the next section in a useful form for EAS development studies
(in the laboratory system of reference). Generally the inelasticity
can be understood as a fraction of interaction energy transferred to
secondary particles created in the interaction. The remaining energy
is carried by the ``leading particle'' and transported downward to
next rank interaction in the hadronic cascade. Thus, the inelasticity
is determined by the mechanism of multiparticle production.
In this paper quantitative results obtained from
particular model calculations will be given.

Initially the geometrical two-chain model (G2C) was developed to describe
relatively low-energy hadron--hadron interactions \cite{g2c}.
Recently
\cite{gmc} the geometrical multichain (GMC) extension of the
model to very high energies was presented. Obtained proton--proton
multiparticle production characteristics show that up to at least
SPS energies the underlying physics can be treated as competitive with
that used by dual parton-like models or relativistic jet (LUND-like)
idea. The straightforward way of the extension of the G2C (and, of course,
GMC) model to proton--{nucleus} and more: to {nucleus}--{nucleus}
interactions was discussed in \cite{nucl,wsrap}.
The possibility of introducing
geometrical chains in the framework of wounded nucleon picture was used
to obtain some results on inelasticity for proton--nucleus collisions.
Comparison with the available data encourage further examinations.

In this paper we would like to present GMC model extension
to {nucleus}--{nucleus} interactions.
The special attention will be paid to chain formations in the
intranuclear cascading process.

The very old and naive picture (simple superposition)
was the first attempt to the problem but it was disqualified on a very
early stage by experiments (see, e.g. \cite{klarhufner}). Then the idea of
wounded nucleons arose and it was generally accepted. It can
be expressed in the statement that once one of projectile nucleons
interacts inelastically with the one from a target nucleus
intermediate states called ``wounded nucleons'' are
created. The spatial extension of the wounded nucleon is the same as
original nucleon before the collision. The subsequent
collisions inside the target nucleus take place before this ``excited state''
hadronize. The non-zero time interval between excitation and hadronization
is confirmed by for example, Bose-Einstein correlation study. Its value
depends on particular definition (it is frame dependent) but it is generally
accepted to be of order of at least 1 fm.

In the wounded nucleon idea for proton--nucleus collisions
each subsequent interaction of the once wounded nucleon leads to the
excitation of one of nucleons from the target. So, in a first
approximation, it adds to the
overall average multiplicity about ${1 \over 2}$ of $p$--$p$ multiplicity
(for respective interaction energy) and leads to the rather small decrease
of the incoming wounded nucleon energy (due to momentum and energy transfer).
This general picture is confirmed very well in many experiments.

Similar results are obtained in DPM-like proton--nucleus pictures were the
first interaction inside the target creates two quark--gluon strings stretched
generally between valence quarks of two colliding hadrons. The 
next interactions each
produce each two additional strings but their quark ends are formed
by the sea quarks on the projectile side thus leading to
$\sim$ $1/2$ of $p$--$p$ multiplicity addition and also to the small change
of the main projectile string energy and momentum.

The situation becomes more complicated when nucleons are involved
on both, target and projectile, sides. In this case there can also appear,
among others, collisions of two wounded nucleons.
The introduction of wounded nucleons on both colliding
nuclei leads to the further decrease of the mean
multiplicity (and inelasticity) per one intranuclear interaction.
Anyhow, this can be solved quite natural by careful investigation
of the so-called
intranuclear cascade. In both, DPM- and LUND-like
interaction models the creation of all chains (strings) can be performed and
is well known (e.g. in FRITIOF realization of the LUND
picture \cite{fritiof} and DTUJETII for DPM \cite{dtujet} ) for some time.

The next step in all models existing so far is to perform the hadronization of
wounded nucleons (chains, strings, excited states). The hadronization as a
typical soft process has to be treated by phenomenological models which
use more or less theoretically justified assumptions. The GMC model
also treats the hadronization in a specific way. Differences
were discussed in \cite{g2c,gmc}. Despite these differences,
results
given by different models are quite similar. Free parameters of all the models
were adjusted to about the same experimental data sample. Anyhow, we do not
expect that the details of the hadronization process descriptions can effect
the general subject of this paper very much, as will be shown later.

After the hadronization process, the created hadrons could, in principle,
appear within one of the colliding nuclei \cite{dtujet}.
Thus, there is a possibility
that they (if their energies are high enough) can further interact inside
the same nucleus. This depends on the freeze-out time --- more
exactly on the distance traversed by the wounded nucleon in the
nucleus rest frame before it hadronizes. Thus, it is expected that when the
interaction energy increases this effect vanishes due to
Lorentz dilatation (even if the freeze-out
time in the wounded nucleon rest system is as small as 1 $\div$ 0.5 fm,
the energy of about 1 TeV/nucleon in laboratory frame, is enough to prevent
it from the hadronizing most of the wounded nucleons within the nucleus).

This is the state of art of the majority of contemporary high-energy hadronic
interaction models.
In this paper we will introduce another sort of intranuclear cascade
process.

The creation of wounded nucleons during the passage of one interacting nucleus
through the other is in general rather well established
\cite{capel}. The cascading of newly created hadrons inside
both nuclei is quite natural. (The importance of this process decreases
with increasing interaction energy due to the finite freeze-out time.)
Interactions between excited states (wounded nucleons, chains, strings) is
not so straightforward, but has been extensively discussed (see e.g.
\cite{wnwn})
as an important contamination in quark--gluon plasma searches and
interpretations high transverse energy events. However, the overall effect
on minimum bias event studies is not very significant \cite{capel}.
For the inelasticity studies the influence is expected to be even less
due to the fact that the energy of both incoming, earlier excited
states has to be conserved in the outgoing wounded states.

However, there is another possibility which we will call hereafter second
step cascading:
an interaction of wounded nucleon from a (target or projectile) nucleus
with another nucleon from the same nucleus before the hadronization occurs.
Wounded nucleons excited to relatively high invariant mass in its nucleus rest
frame of reference move fast due to the momentum and energy
conservation. The effect should intensify with the increasing
interaction energy.
The first reason for this is the increase of the wounded nucleon energy
in the nucleus rest frame. If the energy is high enough the rise of the
nucleon--nucleon cross section starts to play an important role in the
increase of
the probability of subsequent interaction. This probability
also increases due to the increase of the range of the wounded nucleon
within the nucleus (Lorentz dilatation of the freeze-out time).
If this time (or the Lorentz $\gamma$ factor) is high enough almost all wounded
nucleons will abandon the nucleus before secondary hadron creation starts.
Of course there
could always be cases when wounded nucleon hadronize inside nucleus,
for example single-diffraction-type excitations to any small invariant masses.

The great influence of second step cascading process is straightforward.
On the
projectile side (high laboratory energy) the excitation of one initially
untouched nucleons by an excited state going backward in antilaboratory
frame of reference leads to a transfer of a part of nucleon energy
to the secondary produced particles. Their energy in the nucleus 
center-of-mass 
system (CMS) 
are rather small but after transformation to the laboratory system
the effect is expected to be quite considerable.

\section {The GMC model}

The GMC model differs from the other known
interaction models (DPM or LUND-type relativistic string model) in
details of
the treatment of the creation of ``wounded nucleons'' (excited states, chains,
strings). The idea of using structure functions to extract the interacting
valence
quark from colliding hadrons and control colour and/or momentum flow has some
advantages, of course mainly in the reduction of number of free parameters
in a model. It moves the problem of structure function determination
(thus, momenta of quarks) from adjusting them by comparing model
expectations to data with the other kind of physical experiments
outside the soft hadronic scattering domain.
This made the high-energy interaction modelling more reliable and
suited it in a wide context of the high energy-physics, or physics in general.
In the GMC model a phenomenological description of the chain creation is
used with very few parameters to be adjusted directly to the soft
hadronic interaction data.
On one hand, this can be treated as a limitation of the model but, on
the other hand, it can, of course, give a better data description. Some
connections between the GMC model and structure function approach exist, 
and both ways are, in some sense, equivalent.

The geometrization of the interaction picture is in the parametrization of
the multiparticle production process as a function of the impact parameter
of colliding hadrons. As was shown, for example in \cite{g2c},
the elastic scattering
data can lead to quite an accurate determination of a ``matter'' distribution
inside hadrons. To avoid the energy dependence of the hadronic cross section
the hadron sizes are assumed to be scaled \cite{DdD}
according to an energy-dependent factor $r_0$ defined by:

\begin{equation}
\sigma_{\rm inel}\ =\ \pi\: r_0^2   .
\label{sinel}
\end{equation}

The hadron opacity function $\Omega$ is associated with the hadron matter
density
distribution $\rho$ and can be determined precisely from the elastic
scattering data at relatively low energies. If one denotes by $E_0$
the energy for which this determination was performed then the geometrical
scaling can be expressed in the following way

\begin{equation}
\Omega(b) \ = \ \int  {\rm d^2 r} \ \rho _1({\rm \tilde b}) \ \rho _2(
{\rm \tilde b -  r)} 
\label{omega}
\end{equation}

\noindent
where $\tilde b$ is the scaled impact parameter defined as:

\begin{equation}
\tilde b \: = b \: \sqrt{
{
{{\sigma_{\rm inel.}(E_0)} \over {\sigma_{\rm inel.}(E)}}
}
}  .
\label{b}
\end{equation}

The geometrical scaling gives a number of predictions concerning
energetic behaviour of, for example cross section ratios,
multiplicity distributions or differential elastic cross sections.
Some of them are confirmed
experimentally quite strongly such as KNO scaling (for soft components)
whilst others are questioned such as 
$\sigma_{\rm elastic}/\sigma_{\rm total}$.
However, slight modifications of general idea presented above
can restore the agreement with the observations. A very good example is
given in \cite{young}. It is shown that
slow increase of hadronic central opacity
can be responsible for the observed change of elastic-to-total 
cross section ratio and elastic scattering slope change
from ISR to SPS energies. However, 
this does not interfere with our geometrical
approach. As will be shown below, the main properties of hadronization
depend on the ratio of the opacities at given $b$ to the central one.
The change in cross section values for nucleus--nucleus collision
is negligible.

For the multiparticle production processes
it can be expected that more peripheral collisions should lead
to less excited intermediate states. The parametrization which describes
quite well many characteristics seen in experiments
was found \cite{g2c} to be:

\begin{equation}
M_{\rm chain} \ \sim  \ M_0 \: + \:
\left( {{\Omega (b)} \over {\Omega (0)}} \right)
^{\alpha}
\: \left({ {\sqrt{s} \over 2 } \: -  \: M_0 } \right)  
\label{mchain}
\end{equation}

\noindent
where $M_{\rm chain}$ is a chain invariant mass, $\sqrt{s}$ is interaction
CMS
available energy, $M_0$ is a mass of the lightest hadron which can be
formed from the particular quark contents,
and $\Omega$ is defined in the Eq.(\ref{omega}). $\alpha$ is
one of the model parameters found to be equal to 0.48.

After the chain mass creation each chain moves independently according
to energy-momentum conservation law for the time equal to
``freeze-out'' time assumed to be a constant and energy independent
in a chain rest system.
Later the hadronization occurs. Due to the subject of this paper
details are not
very important. They can be found in \cite{gmc} where some results are
given for
$\sqrt{s} \sim$ 20 GeV and SPS energies. In general, the particles are created
uniformly in the phase space with limited transverse momenta taken
from the exponential distribution in $m_\bot^2$. Flavours and spin states
are generated according to commonly accepted rules (with some slight
modifications). For the very high energies (chain masses) the gluon
brehmsstrahlung process can take place leading to the prehadronization
break-ups of the initial chain.

\section{Inelasticity}

The general subject of this paper is the inelasticity behaviour of
nucleus--nucleus interactions. The inelasticity can be defined in many ways.
The definition used here is:

\begin{eqnarray}
K_{\cal NN} =
{{\langle { \rm Energy\ carried\ by\ produced\ secondaries }\rangle} \over
{ \rm Initial\ energy\ of\ the\ projectile\ nucleus}} =
{{E_{\rm sec}^{f}+E_{\rm sec}^{b} } \over
{ E_{\rm in}}} \ =          \nonumber \\
=\ { {(E_{\rm w.n.}^f\ -\ E_{\rm lead.}^f)\ +\
E_{\rm w.n.}^b} \over {E_{\rm in}}} \ = \
{ {E_{\rm w.n.}^f\ k_{{\rm pp}}\ +\ E_{\rm w.n.}^b}
\over {E_{\rm in}}} \ = \ \xi ^f \: k_{{\rm pp}} \: + \: \xi ^b 
\label{knn}
\end{eqnarray}

\noindent
where superscripts $f$ and $b$ denote projectile (forward) and target
(backward) chains, $E_{\rm w.n.}$ is an average energy of wounded nucleons
and $E_{\rm lead.}$ is the mean energy of leading barions created in the final
hadronization of created chains. All energies are given in laboratory
system of reference. The factor $(\: 1\ - \ k_{{\rm pp}}\: )$ denotes an
average fraction of wounded nucleon (chain) energy carried by
the leading barion created from this particular chain.
It is defined as:

\begin{equation}
1 \ - \ k_{{\rm pp}} = {{1} \over {E_{\rm w.n.}^{f}} }
\left( { \ E_{\rm proton}^{f}\ +\ E_{\rm neutron}^{f}
\ - \ E_{\rm anti-proton}^{f}\ - \ E_{\rm anti-neutron}^{f} } \right) .
\label{kpp}
\end{equation}

Values of $k_{{\rm pp}}$ obtained in the framework of the GMC model for
different interaction energies are given in Table \ref{table}.

\begin{table}
\begin{center}

\caption
{The inelasticity in proton--proton interactions in the GMC model
for different energies.}
\label{table}
\vspace{.5cm}

\begin{tabular}{|c|c|c|c|}
\hline
\ $E_{\rm lab}$ [GeV] \ &
\ $\sqrt{s}$ [GeV] \ &
\ $k_{{\rm pp}}$ \  &
\ $K_{{\rm pp}}$ \ \\
\hline
10     & 4.5    & 0.20 & 0.29 \\
$10^2$ & 13.8   & 0.32 & 0.43 \\
$10^3$ & 43.4   & 0.34 & 0.45 \\
$10^4$ & 137.   & 0.36 & 0.46 \\
$10^5$ & 433.   & 0.38 & 0.47 \\
$10^6$ & $1.4\: 10^3$ & 0.39 & 0.48 \\
$10^7$ & $4.3\: 10^3$ & 0.40 & 0.49 \\
$10^8$ & $1.4\: 10^4$ & 0.40 & 0.49 \\
$10^9$ & $4.3\: 10^4$ & 0.40 & 0.49 \\
\hline
\end{tabular}
\end{center}
\end{table}
\vspace{.5cm}


It should be remembered that values of 
$k_{{\rm pp}}$ (third column in Table
\ref{table})
are of course different than the common proton--proton inelasticity defined
as $K_{{\rm pp}} = \left[  E_{\rm lab}  
-  \left( E_{\rm proton} + E_{\rm neutron}
 -  E_{\rm anti-proton} -  E_{\rm anti-neutron}\right) \right] /
{E_{\rm lab}} $
(values are given for a comparison in the fourth column in the Table
\ref{table}).

As one can see in the GMC model, 
the inelasticity of proton--proton $K_{{\rm pp}}$
as well as the 
``chain inelasticity'' $k_{{\rm pp}}$ is almost constant for
energies of interest ---
higher than $\sim$ 100 GeV ($\sqrt{s} \sim$ 15 GeV). This result confirms
the suggestion presented in \cite{DdD2} that a rise of the average
$p_\bot$
and mean multiplicity with energy does not necessary lead to an increase in
the
proportion of energy needed to create the secondary particles. The constancy
of inelasticity is not in contrast to available accelerator data.

The average energy losses in a constant amount of matter traversed
are responsible for a development of nuclear cascade in media.
For a description of a particle passage through the matter, the behaviour
of the inelastic cross section is as important as the inelasticity
in each individual interaction. The rise of the proton--proton
cross section seen in the
high-energy range is well known and measured precisely up to
about $10^{15}$ eV of proton energy in the laboratory system of reference.
It is obvious that this should manifest itself in the proton--nucleus and
nucleus--nucleus cross section behaviour as well.

In the geometrical scaling picture the inelastic proton--proton
cross section is given by:

\begin{equation}
\sigma_{{\rm pp}}^{\rm inel}(b)\ =  1\ - {\rm exp}
(-2\: \Omega(b))
\label{spp}
\end{equation}

\noindent
where $\Omega$ is defined in Eq.(\ref{omega}) and is scaled due to the
cross section increase with energy. The extension to the
nucleus--nucleus
interaction according to the Glauber approach leads to the formula:

\begin{equation}
\sigma_{\cal NN}(b)\ =\ 1 - \int \
\prod _{n=1}^A
{\rm d}^2{b_A}_n
T({b_A}_n)
\prod _{m=1}^B
{\rm d}^2{b_B}_m
T({b_B}_m)
 \left[
{1\: - \:
{\rm d}
\sigma_{{\rm pp}}(\: b-{b_A}_n+{b_B}_m \:)}
\right] 
\label{snn}
\end{equation}

\noindent
where $T(b)$ is a nucleus profile function,
$b$ - an impact parameter of nucleus-nucleus collision and
${b_A}_n$ is a position of the $n$th nucleon of the nucleus $A$
on the impact parameter plane.

\noindent
Further simplifications of Eq.(\ref{snn}) are often used to calculate
nucleus--nucleus cross section. The main is an optical approximation
which treats $\sigma_{{\rm pp}}$ in Eq.(\ref{snn}) as equal to the
$\delta$ function multiplied by respective value of proton--proton
cross section.
Sometimes also an overall opacity of the nucleus as a whole is calculated
by integration of nucleon-nucleon opacity with the nucleus profile
function. Such a treatment leads directly to a result
similar to the one given in Eq.(\ref{spp})
with the respective nucleus opacity instead of $\Omega$ defined originally
for hadronic collision.
As will be shown below, differences between results obtained
with and without such approximations
increase with energy reaching about 30\% at highest energies seen
in cosmic rays. However, it has to be remembered that the extrapolation
of the proton--proton cross-sections so much further from the
region of direct measurements is much more uncertain.
In this paper the
exact numerical integration of Eq.(\ref{snn}) was used as a method for
nucleus--nucleus cross section calculations.

The nucleus profile function is obtained as a projection of
the nucleus density. The assumed density is

\begin{equation}
\varrho (R) \ = \
\left\{
{
\begin{array} {lcc}
{\textstyle {A} \over {\textstyle (a \sqrt{\pi} )^3}}
\ {\rm exp} \ \left[ \: - \: \left( {{\textstyle R} \over {\textstyle a}} \right) ^2 \right]
\ \ \ \ &{\rm for} &\ \ \ A < 40  \\
{\textstyle{ {\rho _0}}
\over
{\textstyle {1 \: + \: {\rm exp} \:[\: ( \:R - c)\: 4.4 / \: t \: ] }}}
\ \ \ \ &{\rm for} &\ \ \ A \geq 40 
\end{array}
}
\right.
\end{equation}

\noindent
which is Gaussian and of Fermi type for lighter and heavier
nuclei, respectively.

The parameters of that formula are fitted to describe the data
on the inelastic cross section. Details are given in \cite{wsrap}.

In most of cross section calculations a
lack of any correlation between nucleons inside the nucleus is assumed.
Of course it is quite obvious that two nucleons can not be very close
to each other. In the present calculations the minimal distance between
each pair of nucleons inside the nucleus was set equal to 0.8 fm.

\begin{figure}
\centerline{\psfig{file=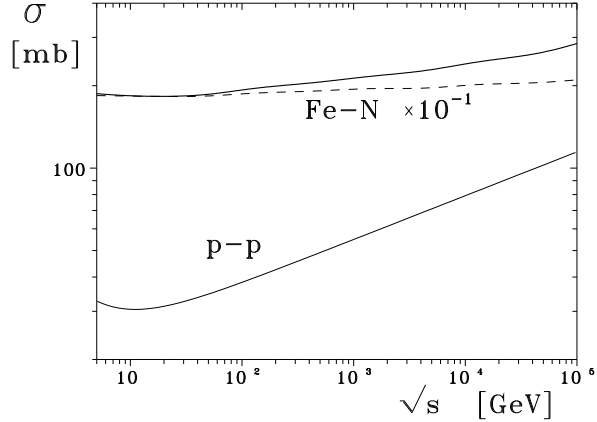,width=9.5cm}}
\caption{Cross section for $p$--$p$ interaction 
used in the present calculations
and the calculated Fe--N inelastic cross section. The broken curve represents
the result obtained using the optical approximation and the full curve 
represents the geometrical scaling.}
\label{fig1}
\end{figure}

Fig.~\ref{fig1} presents energy dependencies of
nucleus--nucleus
cross sections for iron-nitrogen collisions (these
nuclei are chosen because of their importance in the cosmic ray
(EAS) physics) and $p$--$p$ cross sections obtained from
the Regge theory-inspired formula \cite{ppsig}:

\begin{equation}
\sigma_{{\rm pp}}\ =\ 56\ \left( \sqrt{s}\right) ^{\: - \: 1.12}\
+\ 18.16\ \left( \sqrt{s} \right) ^{\: 0.16}
\end{equation}

\noindent
fitted to the available accelerator data.

\begin{figure}
\centerline{\psfig{file=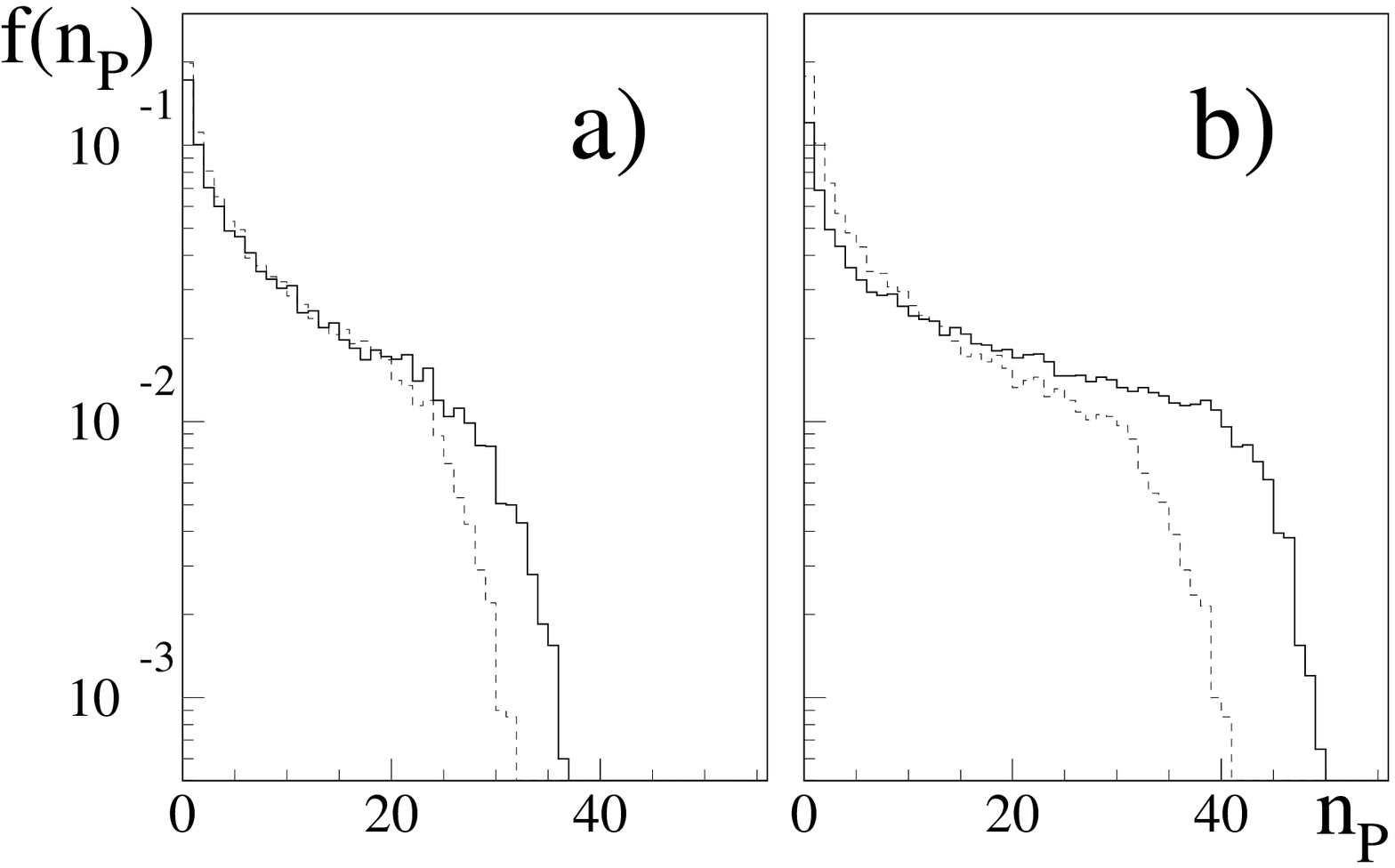,width=9.5cm}}
\caption{Distributions of the number of wounded nucleons in projectile (iron)
nucleus in the Fe--N collision for interaction
energies, $\rm E_{lab}$ is equal to ({\it a}) 
100 GeV and ({\it b}) $10^6$ GeV.
The broken and full curves represent results
without and with second step cascading, respectively.}
\label{fig2}
\end{figure}

The number of nucleons participating in the inelastic nuclei collision
is related closely to the respective cross section ratios
and studied extensively in nucleus--nucleus interaction examinations.
The commonly used approximation based on the optical
approach gives:

\begin{equation}
n_{A}={{A\sigma_{pB}} \over {\sigma_{AB}}}\ \ \ \ \ \ \ \
n_{B}={{B\sigma_{pA}} \over {\sigma_{AB}}}\ \ \ \ \ \ \ \
\nu_{AB}={{AB\sigma_{{\rm pp}}} \over {\sigma_{AB}}}\ \ \ 
\label{nnnu}
\end{equation}

\noindent
where $n_{A}$ in the average number of participants from $A$ nucleus
and $\nu_{AB}$ in the average number of intranuclear inelastic
nucleon--nucleon collisions in the $A$--$B$ interaction.
In the Fig.~\ref{fig2} the distributions of number of
nucleons participating in Fe--N collisions for different interaction energies
are given. Broken curves represent the approximation used in evaluation of
Eq.(\ref{nnnu}). This approach, however, does not take into account the
second step cascading mechanism described qualitatively above.
Thus, these
lines depict only the ``primarily wounded'' nucleons in each colliding
nucleus. This means that they interact inelastically during
the passage of one nucleus through the other (the first step cascading).
After this the second step cascading takes place. The primarily wounded
nucleons inside each nucleus moves back and, when they traverse its own
nucleus, they can excite nucleons which survives the ``primary'' collision
(with nucleons from the other nucleus). Cross sections for the second
step cascading interaction are the same as for $p$--$p$ collision
at respective CMS energy. Thus no additional parameters are needed
to describe the second step cascading process. The effect presented
in Figs.~\ref{fig1} and \ref{fig2} is independent on a particular Monte Carlo
realization. Small differences in $p$--$p$ cross sections are negligible here.

The second step cascading process has to
increase of the number of excited nucleons in both colliding nuclei.
Results of our complete collision calculations are given in the
Fig.~\ref{fig2} by the full histograms.

\begin{figure}
\centerline{\psfig{file=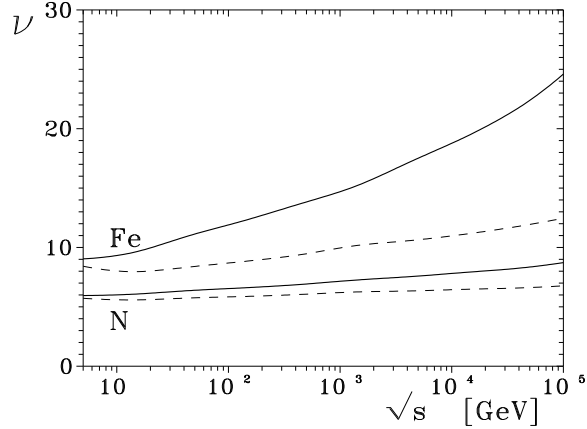,width=9.5cm}}
\caption{Average wounded nucleon number for Fe--N interaction.
The two upper curves are for iron, the lower curve 
for nitrogen nucleus respectively.
Broken curves show the number of primary wounded nucleons
(before second step cascading), full curves represent
final wounded nucleon numbers (after second step cascading process).}
\label{ni}
\end{figure}

The importance of second step cascading for the number of wounded nucleons
is seen even better in the Fig.~\ref{ni} where the mean values of the
wounded nucleon numbers are given as a function of interaction energy.
The effect is stronger on the projectile (heavier) nucleus side.

\begin{figure}
\centerline{\psfig{file=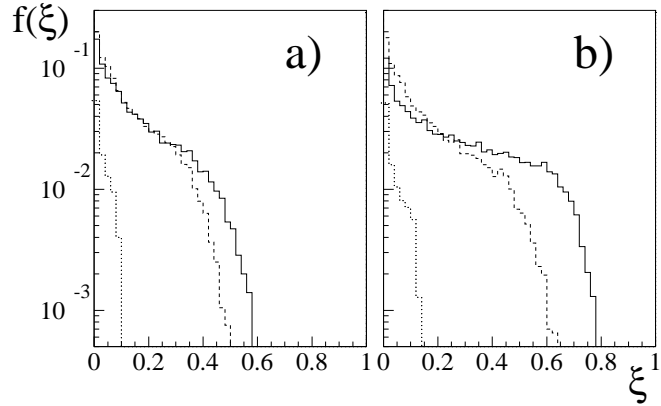,width=9.5cm}}
\caption{Distributions of the energy fraction
carried by wounded nucleons (per 1 nucleon)
in the Fe--N collision for interaction
energies, $\rm E_{lab}$ is equal to ({\it a}) 
100 GeV and ({\it b}) $10^6$ GeV.
The broken and full curves represent results
without and with second step cascading, respectively.
Dotted histograms shows the energy fractions carried
by wounded nucleons from the target (divided by 10).}
\label{fig4}
\end{figure}

\begin{figure}
\centerline{\psfig{file=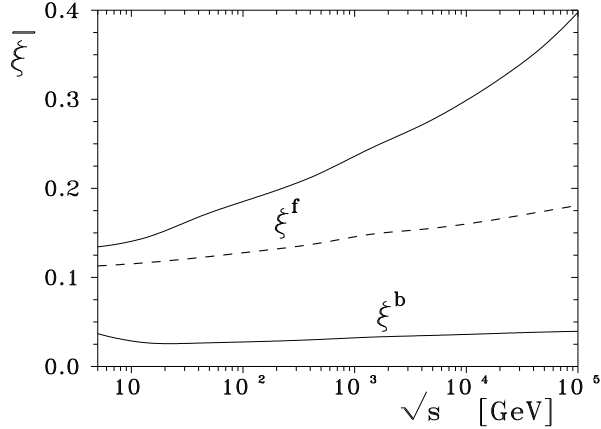,width=9.5cm}}
\caption{Fraction of the laboratory energy carried by wounded nucleons
after the Fe--N collision.
The broken and full curves
represent results
of calculations without and with second step cascading, respectively.}
\label{efr}
\end{figure}

The increase of the number of wounded nucleons has to lead to an 
increase of the
inelasticity no matter how it is defined. To give the quantitative
description of this increase we will use the definition in Eq.(\ref{knn}).
For this the average energy carried by wounded nucleons is needed.
It is obtained from the detailed GMC model Monte Carlo
calculations of nucleus--nucleus
collisions tracking each nucleon in both nuclei during its passage
through the other and after this an eventual traverse through its own
nucleus up to the hadronization phase began.
The respective distributions are presented in Fig.~\ref{fig4} for two
different energies and the respective average values as a function of
interaction energy are shown in the Fig.~\ref{efr}.
Broken curves show the results without second step cascading, full curves
represent our final results.
Dotted histograms gives the energy
fractions carried by CMS backward-going chains --- ``wounded nucleons''
from the target.

\begin{figure}
\centerline{\psfig{file=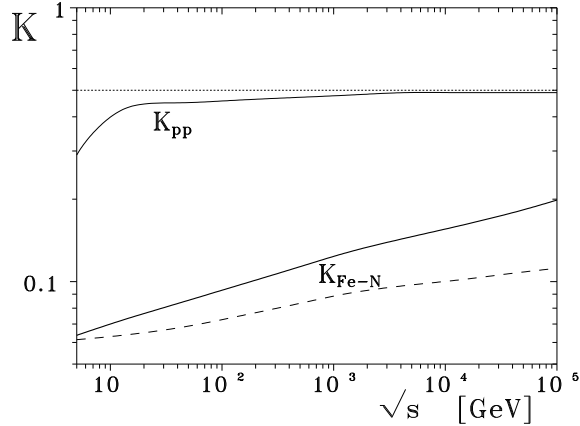,width=9.5cm}}
\caption{Inelasticity coefficient of proton--proton and Fe--N collision
for different interaction energies.
Broken and full curves labeled
by $\rm K_{\rm Fe-N}$ 
were obtained without and with second step cascading
process taken into account, respectively.}
\label{fig6}
\end{figure}

These results together with 
the chain inelasticities ($k_{{\rm pp}}$) listed in
Table \ref{table} can be used to calculate the average inelasticity
in Fe--N collisions ( Eq.(\ref{knn}) ). The final result is presented in 
Fig.~\ref{fig6}.

As can be seen, the introduction of the second step cascading process
leads to a significant increase of inelasticity in nucleus-nucleus
collisions. The excess over the standard cascading calculation results
achieve a factor of 2 at the energies of the order of the highest
seen in cosmic rays (about $10^{20}$ eV).

The review of some interaction properties of different interaction models
(VENUS, QGSM, SIBYLL, HDPM, and DPMJET) is presented in \cite{compa}.
It contains, among others, also values of inelasticity for proton--proton
and Fe--N collisions around interaction laboratory energy 10$^15$ eV.
It is important to note that all the models tested there are introduced
to the CORSIKA \cite{corsika} 
Monte Carlo program for EAS simulations, so they also are
adjusted to cosmic-ray data, while GMC model parameters are obtained
using only accelerator data.
Most of models listed above give $p$--$p$ inelasticities much higher
(0.67, 0.67, 0.64, 0.53, and 0.74, respectively ).
than the one obtained from the GMC model (0.49) at 10$^15$ eV.
In view of the nucleus--nucleus interactions, two models (QGSM and SIBYLL)
give the inelasticity close to the one obtained from GMC model without
second step cascading process (0.093, 0.097 in comparison with .091).
Three other (VENUS, HDPM, and DPMJET) give values close to the one
of GMC with second step cascading (0.14, 0.12, and 0.14 in comparison
with 0.13).
HDPM is the extrapolation of the phenomenological parametrization
of low--energy data (DPM inspired).
The other models are strongly based on the DPM interaction picture.
From the above comparison it is hard to form any conclusive statements.
Difficulties with the reproduction of some interaction characteristics
measured in accelerator experiments by different models presented
also in \cite{compa} additionally confirm such conclusion.

\section{Summary}

From the point of view of the geometrical multichain model
it is quite obvious that the 
intermediate process should take place in between
initial wounded nucleons production and the hadronization.
The importance of this new process is that it effects the number
of wounded nucleons (chains in GMC) leading to an increase of the
inelasticity in {nucleus}--{nucleus} interactions.
The second step cascading in nucleus-nucleus collisions leads to the
significant increase on the inelasticity and wounded nucleon numbers.

This is responsible for
much faster dissipation of the nucleus energy while traversing through
matter. Change, in comparison with the conventional wisdom, of the overall
picture of the hadronic cascades initiated by
nuclei is expected toward a faster development on the first stage
before the nucleus fragment completely. This can be observed
in EAS experiments where
position of shower maxima is higher than expectations based on
many different Monte Carlo
simulation calculations (see, e.g. \cite{flye}). The change of
inelasticity with interaction energy
introduced by second step cascading can also influence the rise of
the EAS maximum
position with primary cosmic-ray particle energy
(see, e.g. \cite{maslo}) which is
also one of the important problems in EAS data interpretation.

\newpage

\end{document}